%% file: paper.tex
\documentclass[runningheads,11pt]{llncs}

\usepackage{tikz}
\usepackage{listings}
\usepackage{amsmath}
\usepackage{amssymb}
\usepackage{mathtools}
\usepackage[hidelinks]{hyperref}
\usepackage{pgfplots}
\usepackage[T1]{fontenc}
\usepackage{stix}

\usepackage{geometry}

\begin{document}

\title{OpenPGP Email Forwarding Via Diverted Elliptic Curve Diffie-Hellman Key Exchanges}

\author{Francisco Vial-Prado\inst{2} \and Aron Wussler\inst{1}}

\authorrunning{F. Vial-Prado \and A. Wussler}
\titlerunning{OpenPGP Forwarding via Diverted ECDH}

\institute {Proton Technologies AG, \email{aron@wussler.it} \and Fortanix, \email{francisco@vialprado.com}}

\maketitle

\input{abstract}

\input{introduction}
\input{scheme}
\input{openpgp}
\input{security}
\input{conclusions}
\input{acknowledgments}

\bibliographystyle{splncs04}
\bibliography{references}

\appendix
\input{appendix-security}
\end{document}

%% file: abstract.tex
\begin{abstract}
 \noindent
 An offline OpenPGP user might want to forward part or all of their
 email messages to third parties. Given that messages are encrypted, this
 requires transforming them into ciphertexts decryptable by the intended
 forwarded parties, while maintaining confidentiality and authentication. It is
 shown in recent lines of work that this can be achieved by means of
 proxy-re-encryption schemes, however, while
 encrypted email forwarding is the most mentioned application of
 proxy-re-encryption, it has not been implemented in the OpenPGP
 context, to the best of our knowledge. In this paper, we adapt the seminal
 technique introduced by Blaze, Bleumer and Strauss in EUROCRYPT'98, allowing
 a Mail Transfer Agent to transform and forward OpenPGP messages without access
 to decryption keys or plaintexts. We also provide implementation details and a
 security analysis.
\end{abstract}

\keywords{Proxy Re-Encryption \and Forwarding \and Elliptic Curve Cryptography \and OpenPGP}

%% file: introduction.tex

\section{Introduction}
Proxy re-encryption is the process of transforming a ciphertext so that it can
be decrypted by a different party than was originally intended.
This transformation is carried out without access
to decryption secrets, plaintexts, or interactive communication with secret-key
holders.
First defined and designed by M. Blaze, G. Bleumer, and M. Strauss
\cite{eurocrypt}, some forms of proxy-re-encryption have already
been applied to ElGamal PGP-encrypted mailing lists \cite{mail-list} in the
context of encrypted email.
In fact, the most highlighted application scenario is email
redirection (see e.g. \cite{improved-pre,pre-lwe,key-private-pre}), however,
the current OpenPGP protocol specification does not include support for this
functionality and, to the best of our knowledge, it has not been implemented in
well-known encrypted email service providers.

A straightforward way to achieve delivery of forwarded encrypted email
is to simply transfer private keys, but there are several drawbacks to
this practice. On one hand, if keys are transferred to
forwarded parties, an attacker controlling them would gain decryption rights
to unforwarded emails, which is not the case in unencrypted forwarding.
On the other hand, private keys could be given to a trusted \textit{Mail
Transfer Agent} (MTA) so that they can decrypt and re-encrypt to forwarded
parties, but this would contradict the trust model of end-to-end encrypted
email services.
The technique we describe aligns with these trust models, proposing to
distribute trust among forwarded parties and the MTA. Security is provided as
long as there is no collusion involving the MTA, i.e. we consider
that the MTA that takes care of the forwarding is a semi-trusted proxy that is
not able to decrypt.

Using two widely used OpenPGP implementations \cite{gopenpgp,openpgpjs}, we
verify the correctness of
the technique in the case where the MTA is also a \textit{Mail
Delivery Agent} (MDA), allowing automatic forwarding between addresses
entrusted to this MDA. However,
in order to enable interoperability with other agents, a
certain \textit{key fingerprint} consideration needs to be ensured by those
services, requiring the modification of a specific OpenPGP packet
to indicate forwarding support.

In this paper, we adapt the technique presented in
\cite{eurocrypt} based on diverted Diffie-Hellman key exchanges, which allows to
emulate proxy re-encryption abilities in the context of symmetrically encrypted
communications following current implementations of the OpenPGP protocol
\cite{RFC4880bis} and involving no additional trust hypotheses, other than
those expected naturally of any email forwarding.
We provide a security analysis and a simple simulation-based proof
of the semi-honest security of the forwarding protocol.

\subsection{Forwarding PGP messages}

The scenario we address is the following:
Bob wants to allow Charles to decrypt email that was originally encrypted to
Bob's public key
without having access to Bob's private key
or any online interaction. Naturally, MTAs should not have the ability to read
the contents of such messages.
We achieve this by proposing a protocol that requires one-time communications
between Bob, Charles, and a trusted MTA:
Bob generates two specific secret elements (a regular secret key, and a
\textit{proxy factor}), securely transfers one to Charles, and the other
to the trusted MTA.

With the proxy factor, the MTA gains the ability to transform any PGP message
encrypted to Bob's public key into another PGP message that can be decrypted
with the newly generated private key, which is now held by Charles. At the same
time, the MTA cannot decrypt the message, nor transform it to another public
key.
The ciphertext transformation is also efficient; upon participating in ECDH key
exchanges, proxies need to store one random field element and two 16-byte key
identifiers per forwarding, and compute a single scalar multiplication on the
elliptic curve per forwarded ciphertext.

\input{fig/forwarding}

\subsection*{A BBS-like transformation in OpenPGP}
\label{subsec-intro-transformation}
The ciphertext transformation technique we describe here is an instance of the
ElGamal-based proxy-re-encryption scheme presented in the seminal paper by
Blaze, Bleumer, and Strauss \cite{eurocrypt}, and as such, is
\textit{bidirectional}, meaning that if a
proxy can transform ciphertexts from Alice to Bob, they also
have the inverse ability.
While this property may be undesirable in most contexts, this is not an issue
in this OpenPGP application. If Bob forwards encrypted email to Charles with
this technique, Charles's forwarded private key is used exclusively to decrypt
Bob's forwarded messages and he has no incentive to receive messages encrypted
to the corresponding forwarded public key, since they could be potentially
forwarded to Bob.

As indicated in \cite{AFGH05}, these schemes are also \textit{transitive},
meaning that if Alice forwards an email to Bob and Bob forwards it to Charles,
nothing prevents proxies from forwarding Alice's emails directly to Charles and
not to Bob.
More generally, any malicious forwarding MTA could compose
re-encryptions and selectively deliver messages to some recipients, ignoring
the rest. While transitiveness appears to be unavoidable here,
such an MTA would be driving against its deliverability incentives and may be
detected by end parties.

The reason why this scheme was chosen over more modern re-encryption schemes is
that an important objective is to maintain backwards compatibility with
incoming messages.
Furthermore, the sender is often oblivious to automatic mail forwarding
happening on the recipient's side, so the chosen scheme must be transparent from
the sender's perspective.
The BBS technique provides a fast and secure scheme that is compatible with
this constraint.

An undesirable property that is partially mitigated by the use of the OpenPGP
protocol, also pointed out in \cite{AFGH05},
is the fact that any collusion between a forwarded party and the proxy results
in the knowledge of the forwarding party's private key.
Fortunately, OpenPGP's structure prevents impersonation attacks: If Charles
above steals Bob's private key with the help of a rogue MTA, he could not sign
messages in the name of Bob or authenticate as Bob.
The forwarding key is derived from an encryption-only sub-key and does not serve
for signing or authentication purposes, because it will not be recognised
as a signing or authenticating key by any compliant OpenPGP implementation.
Nevertheless, in this case, Charles and the rogue MTA can decrypt every message
encrypted to Bob they can obtain, even old or unforwarded.
As a mitigation against collusion attacks we advice to: use short lived keys,
avoid the use of filters, generate new keys upon forwarding set-up, and finally
to deprecate these keys when discontinuing the forwarding.
In other words, we assume that either no collusion with MTAs is possible, or
that forwarded parties are trusted with unforwarded messages.

%% file: fig/forwarding.tex
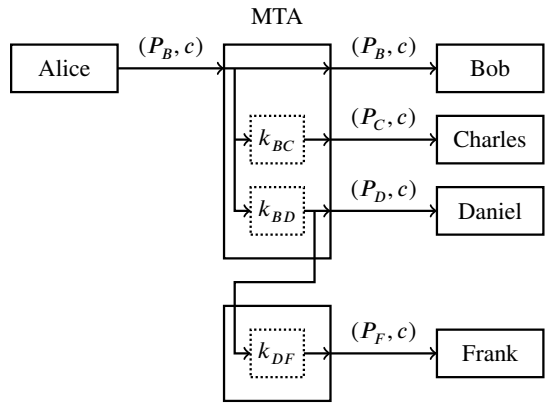
\begin{figure}[ht]
  \centering
  \begin{tikzpicture}[thick, scale=0.7, yscale=0.9]
    \draw[line width=0.3mm] (-5, -0.5) rectangle (-3, 0.5) node[midway] {Alice};
    \draw[line width=0.3mm] (-1, -4) rectangle (1, 0.5) node[midway, yshift=1.8cm] {MTA};

    \draw[->, line width=0.3mm] (-3, 0.0) -- (-1, 0.0) node[above, midway] {$(P_B, c)$};
    \draw[->, line width=0.3mm] (-1, 0.0) -- (1, 0.0);
    \draw[->, line width=0.3mm] (1, 0.0) -- (3, 0.0) node[above, midway] {$(P_B, c)$};
    \draw[line width=0.3mm] (3, -0.5) rectangle (5, 0.5) node[midway] {Bob};

    \draw[->, line width=0.3mm] (-0.8, -1.5) -- (-0.5, -1.5);
    \draw[line width=0.3mm, densely dotted] (-0.5, -2) rectangle (0.5, -1) node[midway] {$k_{BC}$};
    \draw[->, line width=0.3mm] (0.5, -1.5) -- (1, -1.5);
    \draw[->, line width=0.3mm] (1, -1.5) -- (3, -1.5) node[above, midway] {$(P_C, c)$};
    \draw[line width=0.3mm] (3, -2) rectangle (5, -1) node[midway] {Charles};

    \draw[->, line width=0.3mm] (-0.8, 0.0) -- (-0.8, -3.0) -- (-0.5, -3.0);
    \draw[line width=0.3mm, densely dotted] (-0.5, -3.5) rectangle (0.5, -2.5) node[midway] {$k_{BD}$};
    \draw[->, line width=0.3mm] (0.5, -3.0) -- (1, -3.0);
    \draw[->, line width=0.3mm] (1, -3.0) -- (3, -3.0) node[above, midway] {$(P_D, c)$};
    \draw[line width=0.3mm] (3, -3.5) rectangle (5, -2.5) node[midway] {Daniel};

    \draw[->, line width=0.3mm] (0.7, -3) -- (0.7, -4.5) -- (-0.8, -4.5) -- (-0.8, -6) -- (-0.5, -6);
    \draw[line width=0.3mm] (-1, -7) rectangle (1, -5);
    \draw[line width=0.3mm, densely dotted] (-0.5, -6.5) rectangle (0.5, -5.5) node[midway] {$k_{DF}$};
    \draw[->, line width=0.3mm] (0.5, -6) -- (1, -6);
    \draw[->, line width=0.3mm] (1, -6) -- (3, -6) node[above, midway] {$(P_F, c)$};
    \draw[line width=0.3mm] (3, -6.5) rectangle (5, -5.5) node[midway] {Frank};
  \end{tikzpicture}
  \caption{Forwarding scheme -- Alice sends a deferred ECDH ciphertext $(P_B,c)$ to Bob, which the authorized MTA transforms into $(P_C,c), (P_D,c)$ and $(P_F,c)$ using the proxy factors.}
  \label{fig-forwarding}
\end{figure}

%% file: scheme.tex
\section{The forwarding scheme}
While we describe the forwarding technique in terms of large subgroups of
elliptic curves (given the well-known implementation advantages e.g.
\cite{Ber06}), the same technique and security proof can be applied to any
large subgroup where the Discrete Logarithm and Diffie-Hellman problems are
hard, and subgroup membership testing is efficient.
In the remaining sections, let $E$ be an elliptic curve defined over a finite
field $\mathbb{F}$, and $G\in E$ be a generator of a large subgroup of $E$ of
prime order $n$.

\subsection{Asynchronous ECDH Exchanges}
\label{subsec-async-ecdh}

In existing implementations of OpenPGP, asynchronous Elliptic Curve
Diffie-Hellman exchanges are performed to address encrypted email messages (see
figure \ref{fig-ecc}), and work as follows.
Let $d_B \in \mathbb{Z}_n$ be
Bob's private key and $Q_B\coloneqq d_BG\in E$ be Bob's public key.
Alice uses $Q_B$ to generate an ephemeral challenge $P$ and a shared secret
$S \coloneqq d_A d_B G$, that is known only to the two of them, as described in
figure \ref{fig-ecc}.

\input{fig/ecc}

Our purpose is to show that
a selected proxy transmitting a message encrypted to Bob can grant
access to the Alice-Bob shared secret $S$ defined above to selected third
parties, without the proxy knowing $S$ or leaking information about the
forwarding to Alice or other involved proxies.

\subsection{Diverting the secret}
\label{subsec-transformation}
Let us now describe the BBS-like proxy transformation.
Suppose Bob, whose long-term key pair is $(d_B, Q_B)$, wants to forward
incoming messages sent with the protocol defined in \ref{subsec-async-ecdh} to
Charles.
First, Bob generates a new key pair $(d_C, Q_C)$ for Charles and computes the
proxy factor
$$k_{BC} \coloneqq d_B/d_C \mod n$$
for the proxy.

Note that it is not possible to guess $d_B$ from the sole knowledge of
$d_C$ or $k_{BC}$, since the mapping of $\mathbb{F}_n$ onto itself given by
$\phi_y:x\mapsto xy^{-1}$ is bijective for every $y\in \mathbb{F}_n^\ast$ and both
$d_B$ and $d_C$ are sampled uniformly from a subset of $\mathbb{F}_n$ (in
other words, they are indistinguishable from random elements of a subset of
$\mathbb{F}_n$, depending on the chosen curve).
This means that unless the MTA and the forwarded party collude,
they cannot access the secret $d_B$.

Now, given a plaintext $m$, an ephemeral DH shared secret between Alice and Bob
$S = d_Ad_BG$ (also, $S=d_AQ_B$), and a ciphertext $(P_B, c)$ where
$c=\mbox{Enc}_S(m)$ encrypts $m$, the MTA's
objective is to transfer the knowledge of $S$ to Charles, such that
he can decrypt $c$. To achieve this, upon receival of $(P_B, c)$, the MTA
first verifies that $P_B$ is not in a small subgroup of the curve, and then
computes $P_C \coloneqq k_{BC} P_B$.
Why is verifying $P_B$ necessary? Note that $k_{BC}=d_B/d_C$ is not
necessarily a private key of the scheme. In particular, when interpreted as an
integer, $k_{BC}$ may not be a multiple of $h$, the cofactor of the curve, and
therefore the above computation is vulnerable to small subgroup attacks.
Namely, if $P_B$ belongs to a small subgroup of the curve, then $P_C$ is not
necessarily 0, as it would be $d_i P_B$ for any private key $d_i$.
An implementation must abort if $hP_B\neq 0$ (for Curve25519 and Curve448, $h$ is 8
and 4 respectively). According to \cite{RFC7748}, ``\textit{a large number of
existing implementations do not [check the all-zero output]}'', but in this
case, this is mandatory in order to avoid leakage of information about the
proxy factors.

Once $P_B$ is verified, $P_C=k_{BC}P_B$ is computed, the MTA transfers $(P_C,
c)$ to Charles, who in turn computes $d_C P_C = d_C d_C^{-1} d_B P_B = d_A d_B
G = S$, allowing him to decrypt.

In other words, the following equation is computed by involved parties:
\begin{align*}
    S &= d_A d_B G            && \text{(Computed by Alice)} \\
  & = d_B d_A G               && \text{(Computed by Bob)} \\
  & = d_B d_C^{-1} d_C d_A G  && \text{(Computed by Charles)}
\end{align*}
This procedure is described in figure \ref{fig-ecc-forwarding}.
Note that Bob is able to set more than one forwarding address by generating
several valid private keys $d_i$ and corresponding proxy factors.
Given that the forwarding MTA could be selectively forwarding mail to different
users, e.g., using filters, or that the user might want to interrupt the
forwardings at different times, it is compulsory to use unique $d_i$ values for
every different forwarded party.

The scheme is transitive. In fact, Charles can also forward emails further by
repeating the same procedure. Namely, he could generate a new key pair
$(Q_F,d_F)$ that he shares with Frank, and compute $k_{CF}$ to share with the
MTA.

\input{fig/ecc-forwarding}

\subsection{Transformation Proxy as a multiplication oracle}
\label{proof-ecdlp-oracle}
In the procedure described above, Charles could gain access to an arbitrary
stream of forwarded ciphertexts by sending messages to Bob using the same
protocol, and use the forwarded message to obtain information. In this case,
the proxy acts as a multiplication oracle by the secret factor $k_{BC}$:
Charles can choose any $\tilde P\in E$ and any valid ciphertext
$\tilde c$, and submit $(\tilde P,\tilde c)$ to the proxy as a message to Bob.
The proxy verifies and transforms the ciphertext, returning $(k_{BC}\tilde
P,\tilde c)$. Note that guessing the secret factor is exactly solving ECDLP,
thus Charles is not able to obtain any information about $k_{BC}$, or
ultimately $d_B$, since $k_{BC}\tilde P$ is indistinguishable from random. This
is described in figure \ref{fig-ecc-chuck}.

\input{fig/ecc-chuck}

We assume that malicious parties have complete freedom in submitting encrypted
messages to Bob, but note that this activity may be detected by Bob or other
forwarded parties (since they also receive these messages).

%% file: fig/ecc.tex
\begin{figure}[ht]
  \centering
  \begin{tikzpicture}[thick, yscale=0.7]
    \node[align=right, text width=4cm] at (-3, -2) {\bf Alice};
    \node[align=left, text width=4cm] at (3, -2) {\bf Bob};

    \node[align=left, text width=4cm] at (3, -3) {Long-term key pair $(Q_B, d_B)$};

    \draw[<-, line width=0.3mm] (-0.8,-3) -- (0.8,-3) node[midway, above] {$Q_B$};

    \node[align=right, text width=4cm] at (-3, -3) {Retrieves $Q_B$};
    \node[align=right, text width=4cm] at (-3, -3.8) {Secretly picks ephemeral $d_A$};
    \node[align=right, text width=4cm] at (-3, -4.95) {Computes $P = d_A G$, $S = d_A Q_B = d_A d_B G$};

    \node[align=right, text width=4cm] at (-3, -6) {Encrypts $c = \operatorname{Enc}_{S}(m)$};

    \draw[->, line width=0.3mm] (-0.8,-6) -- (0.8,-6) node[midway, above] {$P, c$};

    \node[align=left, text width=4cm] at (3, -6) {Computes $S = d_B P = d_A d_B G$};
    \node[align=left, text width=4cm] at (3, -6.7) {Decrypts $m = \operatorname{Dec}_{S}(c)$};
  \end{tikzpicture}
  \caption{Alice sends an encrypted message to Bob with the Deferred ECDH exchange. $P$: Ephemeral exchange value, ~$S$: Shared secret, ~$m$: Plaintext, ~$(P,c)$: Ciphertext}
  \label{fig-ecc}
\end{figure}
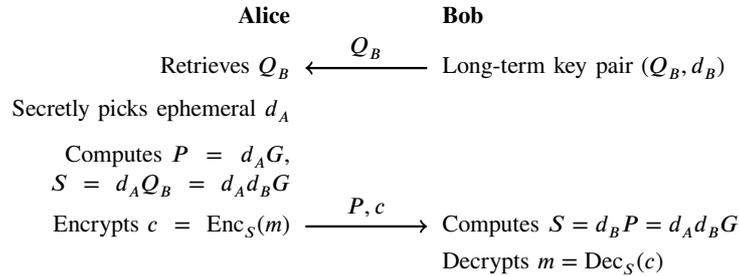

%% file: fig/ecc-forwarding.tex
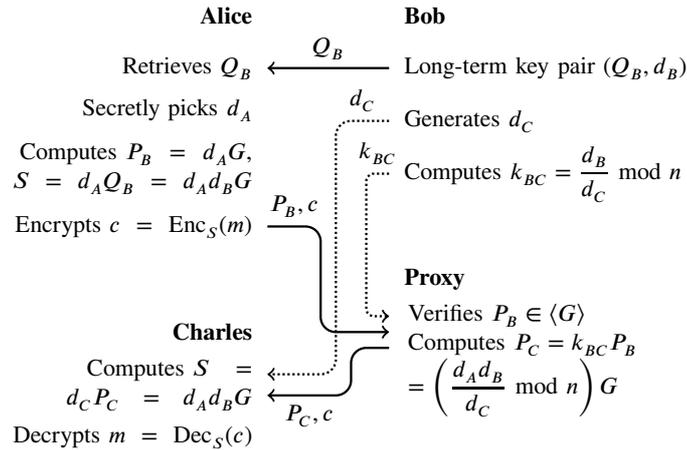
\begin{figure}[ht]
  \centering
  \begin{tikzpicture}[thick, yscale=0.7]
    \node[align=right, text width=4cm] at (-3, -2) {\bf Alice};
    \node[align=left, text width=4cm] at (3, -2) {\bf Bob};

    \node[align=left, text width=4cm] at (3, -3) {Long-term key pair $(Q_B, d_B)$};
    \node[align=left, text width=4cm] at (3, -4) {Generates $d_C$};
    \node[align=left, text width=4cm] at (3, -5) {Computes $k_{BC} = \dfrac{d_B}{d_C}\!\!\mod n $};

    \draw[<-, line width=0.3mm] (-0.8,-3) -- (0.8,-3) node[midway, above] {$Q_B$};
    \draw[->, densely dotted, line width=0.3mm, rounded corners=1mm] (0.8,-5) -- (0.5,-5) node[midway, above] {$k_{BC}$} -- (0.5,-7.7) -- (0.8, -7.7);
    \draw[->, densely dotted, line width=0.3mm, rounded corners=2mm] (0.8,-4) -- (0.1,-4) node[midway, above] {$d_C$} -- (0.1,-8.8) -- (-0.8, -8.8);

    \node[align=right, text width=4cm] at (-3, -3) {Retrieves $Q_B$};
    \node[align=right, text width=4cm] at (-3, -3.8) {Secretly picks $d_A$};
    \node[align=right, text width=4cm] at (-3, -4.9) {Computes $P_B = d_A G$, $S = d_A Q_B = d_A d_B G$};

    \node[align=right, text width=4cm] at (-3, -6) {Encrypts $c = \operatorname{Enc}_{S}(m)$};

    \draw[->, line width=0.3mm, rounded corners=2mm] (-0.8,-6) -- (-0.1,-6) node[midway, above] {$P_B, c$} -- (-0.1,-8) -- (0.8, -8);

    \node[align=left, text width=4cm] at (3, -7) {\bf Proxy};
    \node[align=left, text width=4cm] at (3, -8.5) {$\begin{array}{l}\mbox{Verifies $P_B\in\langle G\rangle$}\\\mbox{Computes }P_C = k_{BC} P_B\\=\left(\dfrac{d_Ad_B}{d_C}\!\!\mod n\right)G\end{array}$};

    \draw[->, line width=0.3mm, rounded corners=2mm] (0.8,-8.3) -- (0.3,-8.3) -- (0.3,-9.2) -- (-0.8, -9.2) node[midway, below] {$P_C, c$};

    \node[align=right, text width=4cm] at (-3, -8) {\bf Charles};
    \node[align=right, text width=4cm] at (-3, -9) {Computes $S = d_C P_C = d_A d_B G$};
    \node[align=right, text width=4cm] at (-3, -10) {Decrypts $m = \operatorname{Dec}_{S}(c)$};
  \end{tikzpicture}
  \caption{Forwarded ECDH exchange. Dotted exchanges are done over an existing secure channel.
    ~$P_B, P_C$: Ephemeral exchange values, ~$m$: Plaintext, ~$c$: Ciphertext}
  \label{fig-ecc-forwarding}
\end{figure}

%% file: fig/ecc-chuck.tex
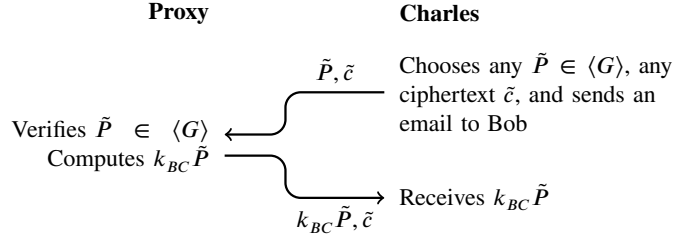
\begin{figure}[ht]
  \hspace*{-1cm}
  \centering
  \begin{tikzpicture}[thick, yscale=0.7]
    \node[align=right, text width=4cm] at (-3.5, -2.5) {\bf Proxy};
    \node[align=left, text width=4cm] at (3, -2.5) {\bf Charles};

    \node[align=left, text width=4cm] at (3, -4) {Chooses any $\tilde P\in \langle G\rangle$, any ciphertext $\tilde c$, and sends an email to Bob};

    \draw[<-, line width=0.3mm, rounded corners=2mm] (-1.3,-4.8) -- (-0.5,-4.8) -- (-0.5,-4) -- (0.8, -4) node[midway, above] {$\tilde P, \tilde c$};

    \node[align=right, text width=4cm] at (-3.5, -5) {Verifies $\tilde P \in \langle G\rangle$ \\ Computes $k_{BC} \tilde P$};

    \draw[->, line width=0.3mm, rounded corners=2mm] (-1.3,-5.2) -- (-0.5,-5.2) -- (-0.5,-6) -- (0.8, -6) node[midway, below] {$k_{BC}\tilde P, \tilde c$};

    \node[align=left, text width=4cm] at (3, -6) {Receives $k_{BC}\tilde P$};
  \end{tikzpicture}
  \caption{The proxy acts as a scalar multiplication oracle when receiving messages from Charles.}
  \label{fig-ecc-chuck}
\end{figure}

%% file: openpgp.tex
\section{OpenPGP implementation}
\label{sec-openpgp}
In this section, we describe some considerations regarding OpenPGP
implementations for parties willing to achieve the encrypted forwarding
procedure described above.

\subsection{Clients tasks}
Assume Bob holds the following long-term keys in order to ensure deferred ECDH
exchanges:
\begin{itemize}
  \item An EdDSA long-term primary key, signing-only;
  \item An ECDH sub-key $(Q_B, d_B)$, encryption-only.
\end{itemize}

\paragraph{Setting up the forwarding}
In order to allow forwarding to Charles, Bob generates another key using the
same curve and parameters as his existing OpenPGP key, with both the EdDSA and
ECDH parts. Let $(Q_C, d_C)$ be the parameters of the encryption-only
ECDH part, that Bob transfers securely to Charles. Finally, Bob computes
$k_{BC}=d_B/d_C\!\!\mod n$ and sends it to the MTA.

\paragraph{Fingerprint selection for KDF} As mandated in the OpenPGP
specification \cite{RFC4880bis}, \S 13.5, a key-derivation function is
called in order to obtain the decryption key for a given message;
in particular, the original
recipient fingerprint is needed as an input to this KDF. Therefore, if
a message was originally encrypted to Bob and forwarded to Charles,
the decrypting implementation needs to use Bob's fingerprint when decrypting a
forwarded message, instead of Charles' fingerprint.
In other words, Bob must specify that the fingerprint
associated to $Q_B$ must be used when decrypting
instead of $Q_C$ in the key-derivation function; otherwise, a fingerprint
mismatch will not allow decryption, since $Q_B$ was used to originally encrypt
this particular message.

This feature prevents tampering with the recipient, but since we need Charles to
decrypt the ciphertext, we propose to change the decryption by altering the field
containing the KDF parameters in the algorithm specific part for ECDH keys.
By adding the original fingerprint and specifying a new version 2, whose field
is already defined and ``reserved for future extensions'', we emulate Bob's
fingerprint in Charles' decryption procedure.
This would not reduce the tampering protection's effectiveness, since this
information is embedded in Charle's key, and at the same time backwards
compatibility for the sender.
Concretely, we propose to alter \textit{variable-length field containing KDF
parameters} defined in \cite{RFC4880bis}, \S 13.5 as follows:
\begin{itemize}
    \item \textit{(Unchanged)} a one-octet size of the following fields; values
        \texttt{0} and \texttt{0xff} are reserved for future extensions;
    \item \textit{(Upgraded)} a one-octet value \texttt{02}, indicating forwarding
        support;
    \item \textit{(Unchanged)} a one-octet hash function ID used with the KD;
    \item \textit{(Unchanged)} a one-octet algorithm ID for the symmetric algorithm used
        to wrap the symmetric key for message encryption;
    \item \textit{(Added)} one-octet of flags with value \texttt{0x01},
        indicating to expect a key fingerprint;
    \item \textit{(Added)} a 20-octet fingerprint to be used in the KDF function; for version
        5 keys, the 20 leftmost bytes of the fingerprint.
\end{itemize}
The forwarded parties' OpenPGP implementation will use this value in the
key wrapping instead of the original fingerprint in order to obtain the right
session key.
Note that with this key pair, only forwarded ciphertexts can be decrypted, and
it does not allow anyone to decrypt messages encrypted to $Q_B$ that were not
transformed by the MTA.

\subsection{Server tasks}
The MTA acts as the re-encryption proxy; it safely stores the factors $k_{ij}$
along with the key IDs to replace them in the ciphertext metadata. When a
matching incoming email arrives, it alters the asymmetric key packet corresponding to the
correct key ID.

An incoming message from Alice to Bob has a \textit{Public-Key Encrypted
Session Key Packets}, that wraps a symmetric key to decrypt the data
packet.  This packet contains:
\begin{itemize}
    \item the curve OID, identifying the correct field;
    \item the ephemeral value $P_B$;
    \item the key ID of Bob's public key;
    \item the encrypted session key.
\end{itemize}

The first step is to parse the ephemeral exchange value $P_B$ and verify that
it belongs to the subgroup generated by $G$
and not other small subgroup of the curve, as mentioned in section
\ref{subsec-transformation}.
This can be achieved by generating any private key $s$ and checking
$sP_B\stackrel{?}{=}0$, or equivalently, checking that $hP_B\stackrel{?}{=}0$,
where $h$ is the cofactor of the curve ($h=8$ for Curve25519). If $P_B$ does
not belong to the large subgroup, the MTA must refuse to process this
ciphertext, as the transformation would leak information about the proxy
factors.

Once $P_B$ is verified, the MTA replaces the ephemeral value of the above
packet with $k_CP_B$.  Also, it replaces the key ID with Charles's key ID to
ensure that his OpenPGP implementation will accept the packet for decryption
(in OpenPGP terminology, the MTA eventually \textit{un-armors} the message,
replaces both values, and \textit{armors} the result).

The MTA can also implement filters based on the unencrypted fields, e.g.,
sender and recipient addresses, or headers. These, of course, rely purely on
trust, e.g., a misbehaving MTA could forward every mail to Charles.

\subsection{Implementation details}

Using the widely-adopted GopenPGP \cite{gopenpgp} and OpenPGP.js
\cite{openpgpjs}, we emulated the forwarding
and verified decryption correctness using those implementation APIs.
Note that the setting up the forwarding is essentially generating two private
scalars, and ciphertext transformation is essentially one scalar multiplication
on the curve, therefore, there is negligible extra cost when supporting this
feature.

Our curve of choice was Curve25519 \cite{Ber06,RFC7748}, defined in the finite
field of $2^{255}-19$ elements, and whose base point $G:x=9$ generates a large
subgroup of prime order
$$n = 2^{252} + 27742317777372353535851937790883648493.$$
Secret keys are sampled randomly from $2^{254}+8\{0,1,2,\dots,2^{251}-1\}$, and
proxy factors are interpreted in $\mathbb{F}_n$ where $n$ is the prime number
displayed above.
The design properties of Curve25519 makes it stand at a comfortable security
level against all known attacks. It is worth noting that scalar multiplication
is implemented in constant time, since double and adding use the same
formul\ae. This consideration is important since any forwarded party can
submit ciphertexts for transformation and measure the time of the
MTA's reaction, with the objective of learning about proxy factors.
Note also that some modern OpenPGP implementations are not implemented in
constant time for other curves.

The verification of small subgroup points in this particular curve is simply
checking if $8P$ is 0, in which case the proxy must refuse to transform the
ciphertext.

%% file: security.tex
\section{Security Analysis}
\label{sec-security}

This section describes how the forwarding protocol is secure against
eavesdroppers, semi-honest
and malicious adversaries, except for collusions between the
proxy and any forwarded party (since they can trivially recover Bob's secret).

A simulation-based proof using the standard techniques from
\cite{Lindell2017} is provided, and works as follows. We first define an
\textit{ideal functionality} $\mathcal F$ associated to the forwarding protocol
$\Pi$. Given a set of participants of $\Pi$, we describe their \textit{views}
and construct \textit{simulators} that produce random corresponding views.
Finally, we show that the simulated views and the output of $\mathcal F$
(i.e., the \textit{ideal world}) are computationally indistinguishable from the
execution of the protocol and its output (the \textit{real world}).

This proof can be found in appendix
\ref{appendix-security}. We provide here an overall analysis of the security
provided by the forwarding protocol.

\subsection{Threat model}
The threat model we consider is an expanded version of the deferred
Diffie-Hellman exchange:
\begin{itemize}
    \item Bob, the original receiver, is always honest: he follows the protocol
        and samples from the correct distributions.
    \item External eavesdroppers, who do not participate but
        may collect all values exchanged in the protocol except for private
        keys and proxy factors (i.e., they know all elements in paths defined
        by bold arrows in figure \ref{fig-ecc-forwarding}).
    \item $F$, the set of forwarded parties, may contain a subset of colluded
        parties that may also send messages to Bob and eavesdrop the protocol.
    \item Alice, the original sender, may collude with forwarded parties.
    \item $T$, the transformation proxy, could misbehave and/or collude with
        other parties.
\end{itemize}
We show that the only collusion that succeeds in attacking is when $T$ colludes
with a forwarded party. This is expected, since they have multiplicative shares
of Bob's secret $d_B$.
To ease notation, for each forwarded party $F_i\in F$, let
$k_i\coloneqq k_{BF_i}$ be the proxy factor held by $T$,
$d_i\coloneqq d_{F_i}$ the secret scalar held by $F_i$ (generated by Bob), and
$P_i\coloneqq P_{F_i}$ the transformed shares (generated by the proxy).

\subsection{Semi-honest parties}
We discuss the semantic security of the execution of the protocol, establishing
that no party can extract information about the plaintext or secrets from the
elements collected throughout the execution, assuming that other parties are
honest (this includes ciphertexts and messages received during the protocol).

\subsubsection{External eavesdroppers}
Let us first discuss security against an adversary that is not participating in
the protocol nor controlling any party, but who intercepts all
communications except for private scalars and proxy factors (these
communications are denoted by dashed paths in figure \ref{fig-ecc-forwarding}).
Additionally, they could send messages to Bob and eavesdrop the
forwarding as described in section \ref{proof-ecdlp-oracle}. Namely, they could
choose any ciphertext $\tilde c$ and a point
$\tilde X$ of the large subgroup and eavesdrop $k_i\tilde X$.
Such a party holds
$$\begin{array}{ll}
    Q_B=d_BG & \mbox{Bob's long term public key,} \\
    P_B=d_AG & \mbox{Alice's DH share to Bob,} \\
    \{P_{F_i}=d_Ak_iG:i=1,\dots,|F|\} & \mbox{set of transformed shares,}\\
    \{k_i \tilde X:i=1,\dots,|F|\} & \mbox{for chosen $\tilde X\in E$,}
\end{array}$$
and is interested in extracting $d_Ad_BG$ (the session secret), any proxy
factor $k_i=d_Bd_i^{-1}$, or any named scalar.
Since proxy factors and secret scalars were generated honestly, note that
this party holds values that are indistinguishable from uniformly
random elements of the large subgroup.
Obtaining any named scalar solves instances of the ECDL problem, and producing
the session secret $d_Ad_BG$ solves the computational ECDH problem.
This proves that the protocol is semantically secure against passive
eavesdroppers.

\subsubsection{Transformation proxy}

The proxy collects the following elements throughout the protocol
$$\begin{array}{ll}
    Q_B=d_BG & \mbox{Bob's long term public key,} \\
    P_B=d_AG & \mbox{Alice's DH share to Bob,} \\
    \{k_i:i=1,\dots,|F|\} & \mbox{proxy factors,}\\
\tilde{X}_1,\tilde{X}_2, \dots& \mbox{shares submitted by other parties (\ref{proof-ecdlp-oracle})}.
\end{array}$$
It is clear that all these elements are uniformly random, provided by honest
parties. Also, again by the ECDH assumption, the proxy alone cannot produce the
session secret $d_Ad_BG$. Naturally, if the proxy and Alice collude (and
Alice is not a forwarded party), they can produce the session secret but cannot
extract Bob's secret $d_B$.
In addition, the proxy cannot compute $d_B$ from the list of proxy factors:
recall that $k_i=d_Bd_i^{-1}\!\!\mod n$, but
since these integers are interpreted in $\mathbb{F}_n$, there is no notion of
GCD (namely, for every $x,y,z\in \mathbb{F}_n$ there exist $u,v$ such that
$x=uz, y=vz$).

\subsubsection{Colluded forwarded parties}

Any forwarded party $F_i$ that also submits a stream of points $\tilde{X}_1,
\tilde{X}_2,\dots$ (as in sec. \ref{proof-ecdlp-oracle}) collects
$$
\begin{array}{ll}
    Q_B=d_BG & \mbox{Bob's long term public key,} \\
    d_i & \mbox{the decryption share,} \\
    P_{i} = d_Ad_Bd_i^{-1}G& \mbox{the transformed DH share,} \\
    k_i\tilde{X}_1,k_i\tilde{X}_2, \dots& \mbox{transformed DH shares of submitted messages}.
\end{array}
$$

Assume further that this party eavesdrop $P_B=d_AG$ from the communication
between Alice and the proxy (or, equivalently, it is colluded with Alice).
Note that, multiplying all DH shares by $d_i$,
this party holds exactly a DH triplet $(d_AG,d_BG,d_Ad_BG)$ and a set of
ECDL samples $(\tilde{X}, d_B\tilde{X})$ for chosen $X$. A collusion between
forwarded parties will only extend the list of ECDL samples, thus,
security follows from both the ECDL and ECDH assumptions.

\subsection{Malicious parties}
\subsubsection{Malicious proxy}
Consider a rogue transformation proxy. It follows from the
ECDH and ECDL assumptions that a malicious proxy cannot extract
secrets $d_A, d_B, d_{F_i}, d_Ad_B$ or $d_Ad_BG$ from its view only. However,
there are other possible misbehaviors of a malicious proxy: restrict the
forwarding to selected parties, alter
proxy factors (sabotaging ciphertexts) or forward messages even if
instructed not to.
We consider these actions as unavoidable for the
forwarding functionality, probably detectable in the context of MTAs, and not
part of our trust assumptions.

\subsubsection{Malicious forwarded parties}
Note that forwarded parties do not control any value in the protocol, with the
exception described in $\ref{proof-ecdlp-oracle}$ where they send messages to
Bob themselves and wait for the proxy's reaction. In this case, they only learn
samples of the form $k_iX$ for chosen $X\in E$ and proxy factors $k_i$ (these are
uniform elements of the large subgroup).
Since this is not a deviation from the protocol, we include these samples in
the semi-honest security proof, therefore,
security against malicious forwarded parties follows.

\subsubsection{Collusion between the proxy and forwarded parties}
Any forwarded party that colludes with the proxy can recover Bob's key:
\begin{equation*}
    d_{i}k_i = d_{i} d_{i}^{-1} d_B = d_B \mod n.
\end{equation*}
We point out that, while recovering this private key may allow to decrypt other
messages, it does not allow to impersonate Bob and generate valid
signatures, since compliant OpenPGP implementations consider this key as encryption-only.

%% file: conclusions.tex
\section{Conclusions}
In this article, we adapted the DH diverting techniques presented in
\cite{eurocrypt} to provide the forwarding functionality to an encrypted email
service compatible with the OpenPGP protocol. This allows encrypted emails to
be securely forwarded with small modifications to an MTA server and the
forwarded parties' clients.
In this context, the scheme we propose
\begin{itemize}
  \item is \textbf{transitive}: Bob can forward his encrypted messages to
      Charles, Charles can forward them to Daniel, and so forth;
  \item is \textbf{non-interactive}: Bob keys' are sufficient to derive the
      proxy transformation factors without any further exchange;
  \item is \textbf{transparent}: Transformed messages are indistinguishable from
      regular PGP messages;
  \item and \textbf{distributes trust}:
      We proved that the only way of gaining access to unforwarded emails is
      for at least one forwarded party to collude with the MTA. In addition, we
      showed that any set of forwarded parties cannot gain access to the
      private key without collusion with the proxy, since they confront several
      independent instances of ECDL and/or ECDH problems.
\end{itemize}
We verified correctness of the forwarded decryption using two well-known
OpenPGP implementations.
While this can already be implemented to allow forwarding within users of the
same mail provider, we described a concrete proposal to the OpenPGP
specification that would imply forwarding compatibility between different
implementations.

%% file: acknowledgments.tex
\section{Acknowledgments}
We thank Ilya Chesnokov, Eduardo Conde and Daniel Huigens for extensive 
discussions, helpful comments and important insights on ECC and OpenPGP.
We would also like to thank Ben Caller for his help with proofreading the paper.

Aron Wussler thanks Daniel Kahn Gillmor for introducing him to proxy re-encryption
schemes for PGP mailing lists.
Also, he wishes to acknowledge the help provided by Prof. Gerhard Dorfer,
who provided valuable help with algebra.

%% file: appendix-security.tex
\section{Simulation-based proof}
\label{appendix-security}
Using the simulation techniques from \cite{Canetti-tutorial,Lindell2017},
we present a security analysis of the forwarding protocol.
This proof addresses the security against semi-honest adversaries, i.e.,
parties that correctly follow the protocol and sample from the correct
distributions, but use all available information to steal secret. Also, a set
of semi-honest adversaries can collude.

The idea of simulation-based proofs is to emulate any set of colluded
participants by a random simulator residing in an ideal world in where a
trusted party exists and the protocol is secure by definition.
The rationale behind this proof
is that if it was possible to extract secrets from the views of colluded
parties, then it would be trivial to distinguish them from the simulator (just
select the view that allow to extract secrets). Conversely, if one cannot
distinguish between the views, colluded parties cannot extract secrets.
Therefore, the objective is to show that the set of elements held by the
semi-honest parties (their \textit{view} of the protocol) are indistinguishable
from the elements held by the simulator.

\subsection{Security definitions}
\label{sec-security-defs}
\subsubsection{Ideal functionalities}
\label{sec-ideal-funcs}
A \textit{functionality} is a process that maps tuples of inputs
to tuples of outputs of a protocol $\Pi$, one for each party involved.
More precisely, for a fixed set $P = \{P_1, \dots, P_k\}$ of $k$
parties participating in the protocol, functionalities are $k$-ary functions
$\mathcal F :(\{0,1\}^\ast)^k\to (\{0,1\}^\ast)^k$ mapping inputs to outputs of
$\Pi$. We write $\mathcal F =(f_1,\dots,f_k)$ where each $f_i$ is a
$k$-ary function that outputs a string.
In addition, if $\mathcal F$ computes the desired outcome by means of a
trusted party in an ideal world that can communicate over perfectly secure
channels with all participants, we say that $\mathcal F$ is an \textit{ideal
functionality}.

\subsubsection{Views}
Given an execution of a protocol $\Pi$ on inputs $\mathcal X=(x_1,\dots, x_k)$,
the \textit{view} of party $P_i$ consists in all elements accessible to $P_i$
throughout the protocol:
$$\mbox{view}_{P_i}(\mathcal X) = (x_i, r_i, m_1^i,\dots, m_j^i),$$
where $x_i$ is $P_i$'s input, $r_i$ is the
content of its internal random tape used to sample elements, and $m_j^i$ is the
$j$-th message it received.
Given a set of colluded parties, their joint view is defined as the tuple
consisting in the concatenation of their views. Also, let
$\mbox{output}_{P_i}(\mathcal{X})$
be the elements held by party $P_i$ identified as the output of the protocol
(note that the inputs or outputs may be empty for some parties).

\subsubsection{Simulators}
Let $\Pi$ be a protocol with inputs $\mathcal{X}=(x_1,\dots,x_k)$. A simulator
is a PPT algorithm that, given an input $x_i$ corresponding to a party of the
protocol, produces a tuple $\mbox{sim}_{i}(x_i)$ similar to the
view of this party. A \textit{joint simulator} takes a set of inputs and
produces a tuple similar to the concatenation of the corresponding parties'
views.

\subsubsection{Simulation-based proof} Following \cite{Lindell2017}, our
notion of security is based on emulating ideal functionalities defined by the
forwarding protocol. This means that, given a protocol $\Pi$ with inputs
$\mathcal{X}$, an ideal
functionality $\mathcal{F}$ computing the output of $\Pi(\mathcal{X})$, and a
set of semi-honest colluded parties, we construct a simulator that takes the
inputs of these parties and produces a random joint view. We then show that
these views along with the output of the protocol (i.e., the \textit{real
world}) and the simulator along with the ideal functionality result of these
parties (i.e., the \textit{ideal world}) are computationally indistinguishable.
For instance, when simulating party $i$, we show that
$$\big(\mbox{view}_{P_i}(\mathcal{X}), \mbox{output}_{P_i}(\mathcal{X})\big)\simeq \big(\mbox{sim}_i(x_i),f_i(\mathcal{X})\big)$$
and similarly for joint views of colluded parties, achieving the proof.

\subsection{Forwarding ideal functionality}
Consider parties Alice, Bob, $T$ (the proxy) and Charlie
(the forwarded party). As before, let $\chi$ be the distribution of
$\mathbb{Z}$
that samples private keys uniformly from a subset of $\mathbb{Z}$,
according to the security requirements of $E$ (for instance, in Curve25519
\cite{Ber06}, private keys are random samples of the form $2^{254}+8m$ for
some $m<2^{252}$).
Assume Alice sends a message $(P_B, c)$ to Bob, and Bob's public key is
$Q_B=d_BG$.

A basic forwarding ideal functionality could be given by
$$\big((P_B,c), (Q_B, k_{BC}, d_C), \cdot, \cdot\big) \mapsto \big(\cdot, \cdot, \cdot, (P_C,c)\big)$$
such that $d_CP_C = d_BP_B$. This works since Charlie receives $d_C$ at some
point in the protocol, and can decrypt $c$ given $S=d_CP_C$. However, note that
this functionality does not
take into account multiple forwarded parties, nor the fact that those
parties can also submit messages to Bob (as in section
\ref{proof-ecdlp-oracle}).
Recall also that, for each forwarded party $F_i\in F$, we note by
$k_i\coloneqq k_{BF_i}$ the proxy factor held by $T$ and
$d_i\coloneqq d_{F_i}$ the secret scalar held by $F_i$.

\begin{definition}
    \label{definition-ideal-func}
    Consider parties Alice, Bob, $T$ (the proxy) and
    $F=\{F_1,\dots, F_m\}$ (the forwarded parties). Let
    $$ \mathcal{X}\coloneqq (\mathcal{X}_A, \mathcal{X}_B,\mathcal{X}_T, \mathcal X_{F_1},\dots, \mathcal X_{F_m}) $$
    be the input of the protocol where:
    \begin{eqnarray*}
        \begin{array}{lll}
            \mathcal{X}_A \coloneqq & P_B, c & \mbox{Alice's message,} \\
        \mathcal{X}_B \coloneqq & Q_B, (k_i, d_i)_{i=1}^m& \mbox{Bob's public key, proxy}\\
                                & & \mbox{factors, and secret shares,} \\
        \mathcal{X}_T \coloneqq & \cdot & \\
        \mathcal{X}_{F_1} \coloneqq & (X_{1j})_{j=1}^{n_1} & \mbox{$F_1$ sends $n_1$ messages to Bob as in \ref{proof-ecdlp-oracle},} \\
            \; \vdots & \\
            \mathcal{X}_{F_m} \coloneqq & (X_{mj})_{j=1}^{n_m} & \mbox{$F_m$ sends $n_m$ messages to Bob as in \ref{proof-ecdlp-oracle}.} \\
        \end{array}
    \end{eqnarray*}
    The forwarding functionality is $\mathcal F = (\mathcal{F}_A,
    \mathcal{F}_B,\mathcal{F}_T,\mathcal{F}_{F_1},\dots, \mathcal{F}_{F_m})$
    where:
    $$
    \begin{array}{ll}
        \mathcal{F}_A : & \mathcal{X} \mapsto \cdot \\
        \mathcal{F}_B : & \mathcal{X} \mapsto c, d_B, P_B, ((X_{ij})_{j=1}^{n_i})_{i=1}^m \\
        \mathcal{F}_T : & \mathcal{X} \mapsto \cdot \\
        \mathcal{F}_{F_1} : & \mathcal{X} \mapsto c, d_1, P_1, ((k_1X_{ij})_{j=1}^{n_i})_{i=1}^m \\
        \; \vdots & \\
        \mathcal{F}_{F_m} : & \mathcal{X} \mapsto c, d_m, P_m, ((k_mX_{ij})_{j=1}^{n_i})_{i=1}^m
    \end{array}
    $$
    such that $d_1P_1 = d_2P_2 = \dots = d_mP_m = d_BP_B$.
\end{definition}
Note that for every message $X_{ij}$ sent as in \ref{proof-ecdlp-oracle}, forwarded
parties also receive an encryption $c_{ij}$ of some message. Without loss of
generality, we omit these encryptions from the ideal functionality, since these
are trivial to simulate and provide no information to attackers (as they are
simply transmitted unchanged throughout the protocol).

More precisely, following section \ref{proof-ecdlp-oracle}, forwarded parties
pick any message $\tilde m$ and a secret key
$\tilde d\leftarrow \chi$, set $\tilde S = \tilde d Q_B$ where $Q_B$ is Bob's
public key, and let $X_{ij} = \tilde d G$, $c_{ij} = \mbox{Enc}_{\tilde S}(m)$.
Instead, without loss of generality, we let forwarded parties freely choose
$X_{ij}$ as an input to the protocol.

\subsection{Security against colluded, eavesdropper semi-honest parties}
\subsubsection{Simulating the semi-honest proxy}
According to the protocol described in section \ref{subsec-transformation},
we have the following view of the proxy throughout the
protocol:
$$\mbox{view}_{T}(\mathcal X) = \big(\cdot, c, P_B, k_1, \dots, k_m, (X_{ij})_{ij}\big)$$
where each $k_i$ was provided by Bob (for the sake of notation, $(X_{ij})_{ij}$
consists in all points chosen by forwarded parties, in definition \ref{definition-ideal-func}). Indeed, for each $i$, Bob sampled $d_i\leftarrow
\chi, k_i \coloneqq d_B/d_i\!\!\mod n$, and sent $k_i$ to the proxy.
Now, consider the simulator that samples $y \leftarrow \chi$, $x_i \leftarrow \chi$, and
$z_i\leftarrow\chi$ for $i=1,\dots,m$, a tuple of random points $(\tilde
X_{ij})_{ij}$ and sets
$$\mbox{sim}_{T}(\mathcal X_T) \coloneqq \big(\cdot, c, yG, x_1z_1^{-1},\dots, x_mz_m^{-1}, (\tilde X_{ij})_{ij}\big).$$
Recall that (i) there is no input or output for $T$ in this protocol, and also (ii) all
other parties behave honestly in this case (in particular, $X_{ij}$ are
uniformly random points of the curve). Given these facts, it is straightforward
to see that view$_T(\mathcal X)$ and sim$_T(\mathcal{X}_T)$ are computationally indistinguishable.

\subsubsection{Simulating forwarded parties}
As described in section \ref{proof-ecdlp-oracle}, each forwarded party $F_i$
computes the session secret $S=d_iP_i$ from the output,
and also has a stream of pairs of the form $(X_{ij},k_iX_{ij})\in E^2$ for
chosen $X_{ij}$ (this is the result of sending messages encrypted to Bob and
parsing the forwarded ciphertexts).
$$(\mbox{view}_{F_i}(\mathcal X),\mbox{output}_{F_i}(\mathcal X)) = \big((X_{ij})_{j=1}^{n_i};c, d_i,P_i,(k_iX_{ij})_{j=1}^{n_i}\big)$$
Note that, since there are no intermediate values computed by forwarded
parties, a simulator that can access the input and output of a forwarded party
can simulate it trivially. In this case, we have simply
$$\big(\mbox{sim}_{F_i}(\mathcal{X}_{F_i}), \mathcal{F}_i(\mathcal{X})\big)\equiv \big(\mbox{view}_{F_i}(\mathcal{X}),\mbox{output}_{F_i}(\mathcal{X})\big).$$
Naturally, the joint view of colluded forwarded parties is also trivially
simulated by the joint simulators.

\subsubsection{Simulating colluded, eavesdropper forwarded parties}
Additionally, let us assume further that this party eavesdropped the share
$P_B=d_AG$ from the communication between Alice and the proxy, and recall that
they also have the public key $Q_B=d_BG$. Such party has the following view and
output:
$$(\mbox{view}_{F_i}(\mathcal X),\mbox{output}_{F_i}(\mathcal X)) = \big((X_{ij})_{j=1}^{n_i};d_AG, d_BG; c,d_i,P_i,(k_iX_{ij})_{j=1}^{n_i}\big)$$
Consider the simulator that samples $x,y\leftarrow \chi$ and sets
$$\big(\mbox{sim}_{F_i}(\mathcal X_{F_i}),\mathcal{F}_i(\mathcal X)\big) = \big((X_{ij})_{j=1}^{n_i};xG, yG;c, d_i,P_i,(k_iX_{ij})_{j=1}^{n_i}\big).$$

The only distinct elements are $d_AG, d_BG$ and $xG, yG$. Note that, since
$d_iP_i=d_Ad_BG$, a DH triplet $(d_AG,d_BG,d_Ad_BG)$ can be
composed in the view. The simulator, on the other hand, can compose the tuple
$(xG, yG, d_Ad_BG)$. However, since $d_Ad_BG$ is uniformly random, this tuple
is indistinguishable from a proper DH triplet $(xG, yG, xyG)$ by the ECDH
assumption.  It follows that the view and the simulator are computationally
indistinguishable. It is straightforward to extend the simulator and address
the case where multiple semi-honest forwarded parties collude: The joint view
and output will only have more independent ECDL samples of the form $(X,
k_iX)$ and parties can compose the same ECDH triplet (also, note that
the additional samples $d_i, P_i$ of the form are also accessible by the
simulator, and the same argument holds).

Putting all these cases together, it follows that the protocol securely
computes $\mathcal F$ in presence of semi-honest, colluded adversaries.